\documentstyle[12pt,epsfig]{article}

\newcommand{\eq}[1]{eq.(\ref{#1})}

\def\gtap{\raisebox{-.55ex}{\rlap{$\sim$}} \raisebox{.4ex}{$>$}}
\def\gsim{\mathrel{\gtap}}

\def\e{\mbox{e}}

\def\G{\mbox{G}}

\begin{document}

%%%%%%%%%%%%% Title
\begin{center}
  {\Large\bf Generation of $10^{15}-10^{17}$ eV photons by UHE CR in  
the Galactic magnetic filed.} \\
  \medskip
  S.~L.~Dubovsky\footnote{E-mail: \verb|sergd@ms2.inr.ac.ru|},
  P.~G.~Tinyakov\footnote{E-mail: \verb|peter@ms2.inr.ac.ru|}\\
  \medskip
  {\small
     Institute for Nuclear Research of
     the Russian Academy of Sciences, 117312 Moscow, Russia  }
\end{center}

\begin{abstract}
We show that the deep expected in the diffuse photon spectrum above
the threshold of $e^+e^-$-pair production, i.e., at energies $10^{15}
- 10^{17}$~eV, may be absent due to the synchrotron radiation by the
electron component of the extragalactic Ultra-High Energy Cosmic Rays
(UHE CR) in the Galactic magnetic filed. The mechanism we propose
requires small ($< 2\times 10^{-12}$~G) extragalactic magnetic fields
and large fraction of photons in the UHE CR. For a typical photon flux
expected in top-down scenarios of UHE CR, the predicted flux in the
region of the deep is close to the existing experimental limit. The
sensitivity of our mechanism to the extragalactic magnetic field may
be used to improve existing bounds on the latter by two orders of
magnitude.
\end{abstract}

%\newpage
\section{Introduction} 

The spectrum of diffuse photons is expected to have a deep of more
than two orders of magnitude at energies $10^{15} - 10^{17}$~eV 
\cite{deep}. This
deep is similar in nature to the well-known Greisen-Zatsepin-Kuzmin
(GZK) cutoff~\cite{GZK} and is caused by electron pair production on
the cosmic microwave background, $\gamma\gamma_b\to e^+e^-$. The cross
section of the latter process reaches its maximum of $0.3\sigma_T\sim
0.2\,\mbox{barn}$ near the threshold at $3\times 10^{14}$ eV and
decreases at higher energies (see, e.g., ref.~\cite{Lee}). The
attenuation length of photons in the region of the deep is of order
100~kpc, so the deep in the spectrum is a universal feature of models
in which high energy photons have extragalactic origin.  Indeed, the
existence of the deep is confirmed by simulations in various models of
UHE CR (for a review see, e.g., refs.\cite{Lee,BhSigl}). For instance,
in the region of the deep the top-down models typically give the
photon flux\footnote{Throughout this paper the ``flux'' denotes the
quantity which is expressed in terms of th edifferential spectrum
$j(E)$ by means of the relation $F=E^2j(E)$.} of order
$10^{-3}$~eV$\,$cm$^{-2}$s$^{-1}$sr$^{-1}$, while at ultra-high
energies, $E\gsim 10^{20}$~eV, the predicted flux is more than two
orders of magnitude larger and reaches $(\mbox{a few})\times
10^{-1}$~eV$\,$cm$^{-2}$s$^{-1}$sr$^{-1}$.

The spectrum of diffuse photons at energies above $\sim 10^{11}$~eV is
known rather poorly. In the region of the deep the bounds have been
obtained by EAS-TOP~\cite{EAS-TOP} and CASA-MIA~\cite{CASA-MIA}.  At
$E\sim 10^{16}$~eV the bound is of order $\sim
0.5$~eV$\,$cm$^{-2}$s$^{-1}$sr$^{-1}$, about two orders of magnitude
higher than has been predicted by top-down models. At ultra-high
energies the photon flux can be as large as $\sim
1$~eV$\,$cm$^{-2}$s$^{-1}$sr$^{-1}$ if the observed UHE CR events are
interpreted as photons (recent results from AGASA \cite{AGASA} suggest
this possibility \cite{Teshima}).

While experimentally the existence of the deep in the photon spectrum
is an open question, theoretically it is not solid
either. Calculations of the spectrum cited above do not take into
account the possibility that high energy photons can be generated in
our Galaxy by UHE electrons via synchrotron radiation in the Galactic
magnetic field. As we argue below, the account for synchrotron
emission may substantially change the photon spectrum at $E\sim
10^{15} - 10^{17}$~eV filling the deep and bringing the expected
photon flux close to the existing experimental limit.

The synchrotron mechanism requires large flux of UHE electrons to hit
the Galactic magnetic field.  It has been recently pointed out in
ref.~\cite{Blasi} that this condition is naturally satisfied in the
halo models of UHE CR (these models explain UHE CR by decays of heavy
relic particles clustered in the Galactic halo \cite{Xparticles}). Due
to the fragmentation process, the decay products of the superheavy
particles contain a large fraction of UHE electrons.

In this paper we show that UHE electrons which are necessary for the
synchrotron mechanism to work can be of extragalactic origin, provided
extragalactic magnetic fields are small. We will see that, in fact,
the large flux of UHE electrons is inherent in top-down models of UHE
CR, so that the generation of high energy photons by the synchrotron
mechanism is a generic prediction of top-down scenarios and is not
specific to halo models of UHE CR .

The key observation is that UHE photons propagate in the extragalactic
space via cascade process being converted to electrons and back with a
small energy loss. As a result, the flux of UHE photons is necessarily
accompanied by the flux of UHE electrons.  At energies of order
$10^{22}-10^{23}$ eV and in the absence of extragalactic magnetic
fields the electron flux is at least as large, or even much larger,
than the photon one\footnote{In the presence of extragalactic magnetic
field the electrons rapidly loose energy via synchrotron radiation and
the argument may not work (see Sect.4 for details).}. While UHE
photons reach the Earth and contribute to the observable flux of UHE
CR, UHE electrons emit synchrotron radiation in the Galactic magnetic
field and transfer their energy to high energy photons. As we will
see below, the energy of produced photons lies in the region of the
deep, while their flux is similar to that of UHE electrons (and, thus,
of UHE photons). Hence, in the absence of extragalactic magnetic
fields the flux of UHE photons at the level of $\sim
1$~eV~cm$^{-2}$s$^{-1}$sr$^{-1}$ implies the flux of synchrotron
photons at the same level which fills the deep in the photon
spectrum. Since the large flux of UHE photons is one of the signatures
of the top-down mechanisms of UHE CR, in the absence of extragalactic
magnetic fields these models generically predict no deep in the
spectrum of diffuse photons.

The mechanism we propose is sensitive to magnetic fields on distances
up to $50$~Mpc from our Galaxy. If no deep in the photon spectrum is
observed and halo models are ruled out, the extragalactic magnetic
field on this distances must be smaller than $2\times
10^{-12}$~G. This is two orders of magnitude better than the existing
bounds \cite{kronberg}.  Inversely, if the deep is found and, at the
same time, UHE CR have a large fraction of photons, the extragalactic
magnetic field must be larger than $2\times 10^{-12}$~G.

The paper is organized as follows. In Sect.2 we estimate the flux of
UHE electrons given the flux of UHE photons and zero extragalactic
magnetic field. In Sect.3 we calculate the spectrum of synchrotron
radiation in the galactic magnetic field for injected electrons of
given energy. In Sect.4 we estimate the effect of extragalactic
magnetic fields. Sect.5 contains our conclusions.

\section{The flux of UHE electrons.}

Our aim in this section is to show that in the energy range
$10^{22}-10^{23}$ eV the flux of UHE photons is necessarily
accompanied by a comparable or larger flux of UHE electrons, provided
the extragalactic magnetic fields are absent. The argument is based on
the observation that at these energies the photon propagation is a
cascade process (see, e.g., \cite{BhSigl}), i.e., propagating photon
is converted to an electron  and back with small energy
loss. As this process is random, one should expect certain ratio of
photons and electrons far from the source.

The main reactions driving the cascade are $e^+e^-$ pair production
(PP) on the radio background, $\gamma\gamma_b \to e^+e^-$, double pair
production (DPP), $\gamma\gamma_b \to e^+e^-e^+e^-$, and the inverse
Compton scattering (ICS), $e\gamma_b \to e \gamma$ \cite{Lee}. In the
energy range of interest double pair production dominates, so one
should expect to find more electrons than photons.

A simple estimate can be obtained if one neglects secondary particles
and energy losses. In this (rather crude) approximation PP and DPP
lead to the conversion of photon to electron with the rates $a_{PP}$
and $a_{DPP}$, respectively, while ICS converts electron back to
photon with the rate $b$. The set of equations which describes
propagation of photons and electrons reads
\begin{eqnarray}
\nonumber
{dn_{\gamma}\over dR} &=& - a n_{\gamma} + b n_e,\\
\nonumber
{dn_e\over dR} &=&  a n_{\gamma} - b n_e,
\end{eqnarray}
where $R$ is the distance from the source, $n_{\gamma}(R)$ and
$n_e(R)$ are fractions of photons and electrons at the distance $R$,
respectively, and $a\equiv a_{PP} + a_{DPP}$. The solution to this system
is
\begin{equation}
{n_e\over n_{\gamma}} = {a\e^{R(a+b)}-C\over b\e^{R(a+b)}+C}, 
\label{ne-over-ng}
\end{equation}
where $C$ is an integration constant whose value is determined by the
ratio $n_e/n_{\gamma}$ at $R=0$.  Far from the source the value of
this constant is irrelevant. 

The observed fluxes $F_{e,\gamma}$ are given by integrals over the
space of $n_{e,\gamma}$ multiplied by the particle injection rate. To
estimate $F_e/F_{\gamma}$ we note that the integrals are dominated by
large distances where the ratio $n_e/n_{\gamma}$ is constant,
\[
{n_e\over n_{\gamma}} \sim {a\over b}. 
\]
Therefore,
\begin{equation}
{F_e \over F_{\gamma}} \sim {a\over b}.  
\label{F/Fnaive}
\end{equation}
Both $a$ and $b$ depend on energy. At $E\sim 10^{22}$ eV one has 
$a\sim 2\times 10^{-2}$ Mpc$^{-1}$, $b\sim 8\times 10^{-3}$
Mpc$^{-1}$ \cite{Lee}, and thus 
\begin{equation}
{F_e\over F_{\gamma}} \sim 2 \quad\quad \mbox{at}\quad 
E= 10^{22}\;\mbox{eV}.
\label{Fe/Fg22}
\end{equation}
At higher energies the rate $a$ is dominated by DPP process and tends
to a constant, $a\to 8\times 10^{-3}$ Mpc$^{-1}$, while the rate $b$
rapidly falls off \cite{Lee}. At $E\sim 10^{23}$ eV \eq{F/Fnaive}
gives
\begin{equation}
{F_e\over F_{\gamma}} \sim 10 \quad\quad \mbox{at}\quad 
E= 10^{23}\;\mbox{eV}.
\label{Fe/Fg23}
\end{equation}

In a more accurate estimate one should take into account energy losses
by leading particles and possible energy sharing in the leading
$e^+e^-$-pair in DPP. In this approximation the result depends on the
energy distribution of initial particles, as well as on the energy
dependence of the rates $a_{PP}$, $a_{DPP}$ and $b$. We have performed
such estimate by dividing energy interval $10^{21} - 10^{24}$ eV into
10 energy bands and solving numerically the system of 20 coupled
equations analogous to eqs.(\ref{ne-over-ng}). We have found that the
corrections to eqs.(\ref{Fe/Fg22}) and (\ref{Fe/Fg23}) are small and
do not change our conclusions, unless the extragalactic magnetic filed
is non-zero (the latter case is considered in Sect.4).

\section{Synchrotron radiation in the galactic magnetic field.}

Consider now the synchrotron radiation of UHE electrons in the
Galactic magnetic field. An ultra-relativistic particle of energy $E$
moving in the magnetic field $B$ emits radiation at the characteristic
frequency \cite{LL}
\begin{equation}
\omega_c = {3\sqrt{\alpha} B\over 2 m_e^3} E^2 
= 6.7\times 10^{14} \left({E\over 10^{20}\,\mbox{eV}}\right)^2
\left({B\over 10^{-6}\,\mbox{G} } \right) \,\mbox{eV}. 
\label{omega-sync}
\end{equation}
The width of the frequency band is roughly $\delta\omega\sim
\omega_c$. As a result of this process, the particle looses energy at
the rate 
\begin{equation}
{dE\over dx} = - {2\alpha^2B^2\over 3m^4} E^2.
\label{energy-loss-eq}
\end{equation}
Both equations are written for the case of particle momentum normal to
the direction of the magnetic field. Generalization to other cases is
straightforward.

Eq.(\ref{omega-sync}) implies that in the Galactic magnetic field,
$B\sim 10^{-6}$~G, electrons with energy $E\sim 10^{20}$~eV
radiate at the characteristic frequency $\omega_c \sim
10^{15}$~eV. This process is the source of high energy photons in the
Galactic halo models of UHE CR \cite{Blasi}.  Since the Galactic
magnetic field far from the Galactic center is smaller, the
extragalactic electrons which we consider in this paper should have
higher energy in order to produce synchrotron radiation in the same
frequency range.

For the quantitative analysis of the synchrotron emission by the
extragalactic electrons consider an UHE electron moving in a varying
magnetic field $B(x)$ perpendicular to its velocity. Integration of
\eq{energy-loss-eq} gives
\begin{equation}
{1\over E(x)}-{1\over E_0} = {2\alpha^2 \over 3 m_e^4} \int_{-\infty}^x
B^2(x) dx,
\label{E=E[x]}
\end{equation}
where $E_0$ is the initial energy of the electron. 

\begin{figure}[ht]
\begin{center}
\epsfig{file=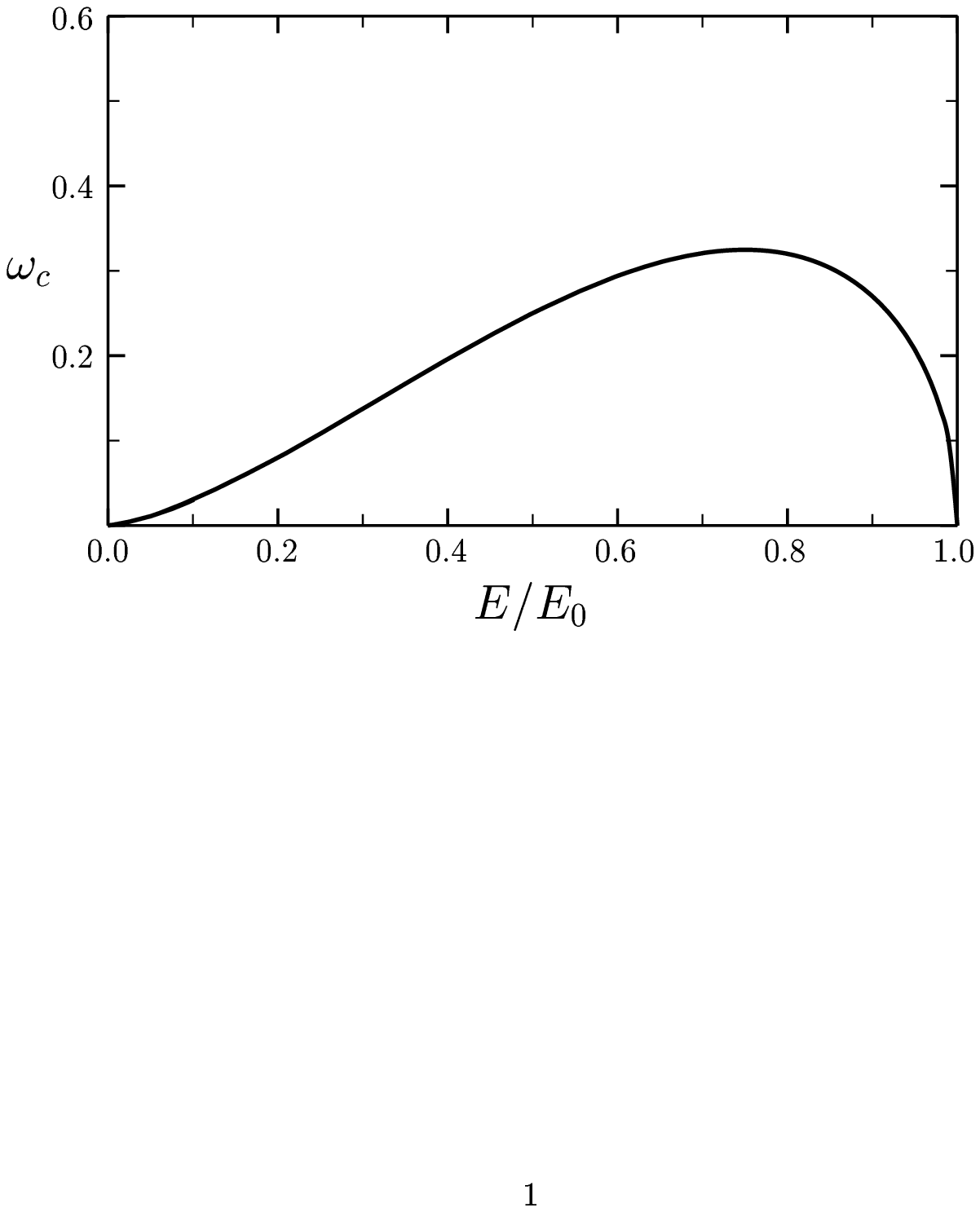,%
bbllx=100pt,bblly=350pt,%
bburx=470pt,bbury=570pt,
%width=236pt,height=175pt,%
clip=}
\end{center}
\caption{The dependence of $\omega_c$ (arbitrary units) on particle
energy for the exponential model of the Galactic magnetic field.}
\end{figure}
For definiteness, let us take the exponentially decaying magnetic
field, 
\begin{equation}
B(x) = B_0\exp(x/x_0)
\label{B}
\end{equation}
(note that we consider a particle propagating from $x=-\infty$). This
behavior is expected in some recent Galactic magnetic field models
\cite{Stanev} for the field in the direction normal to the Galactic
disk. The scale $x_0$ is of order 4~kpc. Making use of
eqs.(\ref{E=E[x]}) and (\ref{B}) one finds the relation between $E$
and $B$ at a given point of particle trajectory,
\[
{1\over E}-{1\over E_0} = {\alpha^2 x_0\over 3 m_e^4} B^2. 
\]
This equation, together with \eq{omega-sync}, determines the dominant
radiation frequency as a function of particle energy, 
\[
\omega_c(E) = {9 E_0^{3/2}\over 2 m_e \sqrt{3\alpha x_0}} f(E/E_0),
\]
where 
\[
f(y) = y^{3/2}\sqrt{1-y}. 
\]
This function is shown in Fig.1. As can be seen from the picture, most
part of the electron energy is emitted at frequencies close to
\begin{equation}
\omega_{max} = \omega_c(3E_0/4) = 
{27 E_0^{3/2}\over 32 m_e \sqrt{\alpha x_0}}
= 0.8\times 10^{15}\left( {E_0\over 10^{22}\,\mbox{eV} }\right)^{3/2}
\left( {x_0\over 4\;\mbox{kpc}} \right)^{-1/2} \mbox{eV}
\label{omegamax}
\end{equation}
According to eqs.(\ref{omega-sync}) and (\ref{omegamax}), at
$E=10^{22}$~eV the electron looses most part of its energy in the
region where the magnetic field is $\sim 10^{-10}$~G, i.e., at the
distance $\sim 36$~kpc from the galactic disk for $B_0 \sim
10^{-6}$~G.

In the case of magnetic field not perpendicular to particle velocity,
the spectrum of synchrotron photons is softer. The same is true for
the magnetic field which falls off slower than in \eq{B}, as usually
assumed for the Galactic magnetic filed in the direction parallel to
the Galactic plane. Thus, one should expect angular dependence of
photon spectrum with more energetic photons coming from the direction
normal to the Galactic plane, and softer spectrum from the directions
in the Galactic plane.

Finally, let us estimate the flux of synchrotron photons assuming the
flux of UHE photons which is typical for top-down scenarios, $(\mbox{a
few})\times 10^{-1}$~eV$\,$cm$^{-2}$s$^{-1}$sr$^{-1}$ at energies
$\sim 10^{22}-10^{23}$~eV. Eq.(\ref{Fe/Fg22}) implies that, outside
the Galactic magnetic field, there is at least as large flux of UHE
electrons which transfer their energy to high energy photons in the
Galactic magnetic field. Since the synchrotron spectrum has
$\delta\omega\sim \omega$, the energy conservation implies that the
flux of synchrotron photons is approximately the same as the flux of
UHE electrons, which is larger by the factor $F_e/F_{\gamma}$ than
the flux of UHE photons.

\section{Effect of extragalactic magnetic field}

As it was shown above, in the absence of extragalactic magnetic fields
the observed flux of $10^{15}-10^{17}$~eV photons is proportional to
the flux of UHE photons and the ratio $F_e/ F_{\gamma}$ at energies
above $E\gsim 10^{22}$ eV.  The presence of large enough extragalactic
magnetic field can significantly decrease this ratio. Indeed, if
$\gamma\to e$ conversion length, $a^{-1}=(a_{PP}+a_{DPP})^{-1}$, is
large compared to the energy loss length of the electron due to the
synchrotron radiation in the extragalactic magnetic field, the flux of UHE
electrons should be much smaller than the flux of UHE photons.

Let us estimate the value of the extragalactic magnetic field at which
the ratio $F_e/F_{\gamma}\sim 1$ at $E\gsim 10^{22}$ eV. For this
purpose note that the solution to eq.(\ref{energy-loss-eq}) in the
constant magnetic field $B$ can be written in the form
\begin{equation}
\label{constfield}
l={3m^4\over 2\alpha^2B^2E}\left(1-{E\over E_0}\right)\;,
\end{equation}
where $l$ is the distance passed by the electron while its energy
decreases from $E_0$ to $E$. Eq.(\ref{constfield}) implies that
electrons with energy $E$ can only come from distances smaller than
\begin{equation}
l_E={3m^4\over 2\alpha^2B^2E} \sim 50\left({2\times 10^{-12}\G\over
B}\right)^2 \left({10^{22}\mbox{eV} \over E}\right) \mbox{Mpc}\;.
\end{equation}
Taking into account that the length $a^{-1}$ of $e\to\gamma$
conversion is of order 50 Mpc at $E=10^{22}$ eV \cite{Lee} one finds
that the extragalactic magnetic field should be smaller than $2\times
10^{-12}$~G at the distances $\leq 50$ Mpc from our Galaxy in order
that the ratio $F_e/F_{\gamma}$ to be comparable or larger than one.
If the magnetic field at distances of order 50~Mpc is noticeably
larger than $2\times 10^{-12}$~G, the ratio $F_e/F_{\gamma}$ is much
smaller than one and the flux of UHE electrons is not sufficient to
fill the deep in photon spectrum by synchrotron mechanism (for the
discussion of present limits on the extragalactic magnetic field see,
e.g., ref. \cite{EGMF}).

It is worth noting that the mechanism we propose is not sensitive to
magnetic fields at distances larger than $\sim 50$~Mpc because this
distance is sufficient for generation of a large fraction of
electrons, $F_e/F_{\gamma}\sim 1$. The effect of distant magnetic
fields is mere decreasing of the UHE photon flux, which is not
important for our argument since we normalize UHE photon flux to the
observed flux of UHE CR.

\section{Conclusions}

To summarize, we have shown that the expected deep in the photon
spectrum at energies $10^{15}-10^{17}$~eV may be absent due to the 
synchrotron radiation of UHE CR in the Galactic magnetic field. The
mechanism we propose involves extragalactic UHE electrons produced
during cascade propagation of UHE photons and thus requires small
extragalactic magnetic fields and large fraction of photons in UHE CR
at energies $10^{22}-10^{23}$~eV.  The latter is the characteristic
feature of all top-down models of UHE CR. Note, however, that our
mechanism does not rely on any peculiarities of these models.

When synchrotron radiation in the Galactic magnetic field is taken
into account, the top-down models predict the flux of
$10^{15}-10^{17}$~eV photons which is close to the current
experimental limits, provided that extragalactic magnetic fields are
small. Thus, it is important to improve the sensitivity of the
experiments in the energy range $10^{15}-10^{17}$~eV. The detection of
the diffuse photon flux at the level of $\sim
10^{-1}$~eV$\,$cm$^{-2}$s$^{-1}$sr$^{-1}$ would strongly suggest that
UHE CR are produced by a top-down mechanism. Moreover, this would
imply that either the mechanism based on the halo model \cite{Blasi}
or the one proposed here works. The two possibilities can be
distinguished by measuring the angular anisotropy of UHE CR \cite{DT}
and of produced high energy photons \cite{Blasi}. Additional signature
of our mechanism is the dependence of the photon spectrum on the
arrival direction as discussed in Sect.3.  On the contrary, if the
photon flux in the region of the deep is smaller than $\sim
10^{-3}$~eV$\,$cm$^{-2}$s$^{-1}$sr$^{-1}$ and, at the same time, UHE
CR have a large fraction of photons, the extragalactic magnetic field
must be larger than $2\times 10^{-12}$~G.

\section*{Acknowledgments}

The authors are indebted to D.S.~Gorbunov, O.E.~Kalashev,
M.V.~Libanov, V.A.~Rubakov and D.V.~Semikoz for useful discussions and
comments.  The work of S.D. is supported in part by Russian Foundation
for Basic Research under grant 99-02-18410, INTAS grant 96-0457 within
the research program of the International Center for Fundamental
Physics in Moscow, and ISSEP fellowship.


\begin{thebibliography}{99}
\bibitem{deep} R.J. Could, G. Schreder, Phys. Rev. Lett. {\bf 16}
(1966) 252;\\
J.P. Jelley,  Phys. Rev. Lett. {\bf 16} (1966) 479.
\bibitem{GZK} K. Greisen, Phys. Rev. Lett. \textbf{16} (1966) 748;\\
G.T. Zatsepin and V.A. Kuzmin, Pisma Zh. Eksp. Teor. Fiz. \textbf{4}
(1966) 144;
\bibitem{Lee} S. Lee, Phys. Rev. \textbf{D58} (1998) 043004;
\bibitem{BhSigl} P. Bhattacharjee, G. Sigl, {\em Origin and
Propagation of Extremely High Energy Cosmic Rays}, astro-ph/9811011;
\bibitem{EAS-TOP} M. Aglietta \emph{ et al.},
Astropart. Phys. \textbf{6} (1996) 71;
\bibitem{CASA-MIA} M.C. Chantel \emph{ et al.},
Phys. Rev. Lett.\textbf{79} (1997) 1805;
\bibitem{AGASA} M.~Takeda {\it et al.}, Phys. Rev. Lett. {\bf 81}
(1998) 1163, astro-ph/9807193.
\bibitem{Teshima} M. Teshima, talk presented at 10-th International
School ``Particles and Cosmology'', Baksan, 1999.
\bibitem{Blasi} P. Blasi, {\em Gamma Rays from Superheavy Relic
Particles in the Halo}, astro-ph/9901390;
\bibitem{Xparticles} V.A. Kuzmin and V.A. Rubakov,
Phys. Atom. Nucl. \textbf{61} (1998) 1028; V. Berezinsky,
M. Kachelriess and A. Vilenkin, Phys. Rev. Lett. \textbf{79} (1997)
4302;
\bibitem{kronberg} P.P. Kronberg, Rep. Prog. Phys., \textbf{47} (1994)
325.
\bibitem{LL} L.D.~Landau and E.M.~Lifshits,``Field Theory'', Moscow, 1973.
\bibitem{Stanev} T.~Stanev, Astrophys. J., {\bf 479} (1997) 290.
\bibitem{EGMF} D. Ryu, H. Kang, P.L. Biermann, A\&A, {\bf 335} (1998)
19, astro-ph/9803275;\\
T.A. En{\ss}lin, P.L. Biermann, P.P. Kronberg, and X.P. Wu; Astrophysical
Journal {\bf 477} (1997) 560, astro-ph/9609190;
\bibitem{DT} S.L. Dubovsky, P.G. Tinyakov, Pisma Zh. Eksp. Teor. Fiz.,
{\bf 68} (1998) 99.
\end{thebibliography}
\end{document}